\documentclass[11pt]{article}
\usepackage{geometry}                % See geometry.pdf to learn the layout options. There are lots.
\usepackage{graphicx}
\usepackage{amssymb}
\usepackage{epstopdf}
\usepackage{amsmath}
\usepackage{fixmath}
\usepackage{MnSymbol}
\usepackage{stmaryrd}
\usepackage{amsfonts}
\usepackage{array}
\usepackage{multirow}
\DeclareMathAlphabet{\mathpzc}{OT1}{pzc}{m}{it}

\title{A Set of Statistical Variables for Hydrodynamic Flow}
\author{Clifford Chafin\\ \small{Department of Physics, North Carolina State University, Raleigh, NC 27695} \thanks{cechafin@ncsu.edu}}

\begin{document}
\maketitle

\begin{abstract}
Through a discussion of some typical unsteady hydrodynamic flows, we argue that the time averaged hydrodynamic functions at each point give a rather sparse filling of the local jet space.  This situation then suggests a set of time dependent probability functions that are shown to give evolution uniquely defined by the Navier-Stokes equations through a set of ``differential distribution equations.''  The closure relations are therefore unique and have no ad hoc characteristics.  Annealing methods are proposed as a way to arrive at the stable stationary solutions corresponding to time averaged fluid flow with constant driving forces and fixed boundary conditions.  Some applications of this method to quantum statistical mechanics and kinetic theory to higher orders are suggested.
\end{abstract}

The treatment of turbulence and otherwise unsteady flow in hydrodynamics is an old problem with enduring relevance \cite{Batchelor}.  Given that most large energetic fluid systems have enormous Reynolds numbers, turbulence is an essential part of the process of conveying strain, heat and material across them.  Even as a microscopic physical understanding of fluidity and the liquid state remains elusive, so fluids have enduring mysteries at large scales.  Computational power for simulation grows while analytic approaches and closure schemes for them proliferate yet our confidence in the results is often undermined by their ad hoc nature.  

Anticipating that turbulence would be adequately characterized by a mean flow and some fluctuations about it, Reynolds introduced an averaging scheme for the velocity $u=\bar{u}+u'$.   This gives a mean motion driven by the mean pressure and mean viscosity but with a density dependent ``stress'' term \cite{Batchelor}.  How to treat this stress term is the main complication in the analysis of hydrodynamics of turbulent flow.  Analogously, in kinetic theory, the collision term in the evolution of the distribution function $f(x,v,t)$ is handled by the assumption of ``molecular chaos.''  In the case of kinetics, this works very well to low orders in the flow gradients and there are quantitive bounds on why this should be so \cite{Villani}.  For fluid dynamics we do not have such a powerful condition suggesting the problem is either much harder or we have not chosen the best variables.  

In this article we will introduce a new set of variables to give a statistical analysis of unsteady flow that is analytical in nature.  Specifically, we introduce a probability function, normalized to one at each point, $f(x,v,\partial v,\partial\partial v,\ldots, \rho, \partial\rho, \partial\partial\rho,\ldots)$ corresponding to the long time average of the local flow and its correlations.  Of course, truncation will be necessary as will justifications for why and where this is to be done.  Most importantly, we will be interested in stationary boundary conditions and the resulting flows.  

Since ensembles play a standard role in the derivation of hydrodynamics from statistical mechanics and we are interested in a statistical treatment of flows, it is appropriate to give a brief discussion on the relevance and limitation of ensembles to physics in general.  Classically, a gas is a particular set of particle positions and motions that we can view as a single point in phase space.  The conserved quantities of energy, momentum\ldots constrain the gas to move about in a reduced dimensional foliated subset of this manifold.  

Although accurate, this does not give a very useful description of a gas.  We can only measure large scale averaged quantities like pressure, density, velocity and such so we could imagine considering an equivalence class of all such states and evolving.  Fluctuations and small scale variations can grow and lead to large scale deviations so this method has its limitations but, the implicit assumption of statistical mechanics is that this not a problem for most purposes.  For static (i.e.\ flow free) gases this is valid on long time scales set by entropy considerations.  For dynamic flows, especially driven ones, this is not nearly as good an assumption.  

A second consideration is the validity of 
scales of parcel averaging.  One can average over finite volumes at one time but then this places limits on the size of flow variation we can consider.  Alternately, one can average over smaller volumes over finite times but this places other constraints on validity as flows change and such a consideration is limited to lagrangian parcel averages.  
Subcritical fluctuations can exist and these can grow to give large scale differences in turbulent flows.  Even if we are only interested in local fluctuations, these can give important contributions to macroscopic observables like the viscosity \cite{Romatschke}.  It is hard to see how the kinetic ensemble distribution functions can capture this reality.

\section{Aspects of Statistical Flow}
Laminar flow is the case of a flow that persists in a stationary fashion at each point.  Such flow is either static or found in the case of uniformly driven boundary conditions or external forces.  The well-definedness of the term is a bit in doubt in the case of time varying boundaries.  If the boundaries move in a periodic fashion and the flow does as well, it might be considered to be laminar even though the flowlines are not time independent.  An unambiguous notion of unsteady flow might be given by the notion of bifurcation as we see in the case of the van Karman vortex street.  These flows around obstacles give a set of counterrotating vortices that grow and advance behind the body.  The reason we call this a bifurcation is that, in the Eulerian point of view, the flow velocity at each point undergoes a simple period of evolution at each point.  

Given the values of $v(p_{o})$ for a particular point $p_{o}$, we can find the neighboring $v$ values from the function $\partial v= s(p_{o},v(p_{o},v(p_{o})))$ by integration.\footnote{Note that the choices of the functions $s$ are not arbitrary but must satisfy the Frobinius integration condition $s_{ij}=s_{ji}$.}  The pressure can be found from the the equation of state after integrating the density $\rho(p_{o})$ with $\partial\rho=u(p_{o},v(p_{o},v(p_{o})))$.  This particular case gives a foliation of solutions $\{v(p),\rho(p)\}$ indexed by the one parameter set of value of $v$ at $p_{o}$. The Navier-Stokes equations also require second derivative information in the velocity $\partial\partial v=r(p_{o},v(p_{o},v(p_{o})))$ To relate this to the N-S equations requires us to give a time derivative of each one of these functions that preserves the foliation and advances through it at a rate indicated by the relative density $h$ at each point.  Let these solutions be indicated by $v(p,\lambda)$, $\rho(p,\lambda)$ where $\lambda$ is a compact parameter.  

We can write the N-S equations, assuming that the fluid is incompressible enough during its motions that bulk viscous contributions are negligible, as
\begin{align}
\partial_{t}v+v\cdot\nabla v=-\nabla P+\eta\nabla^{2}v
\end{align}
Our parameterized set of solutions then evolves along the parameter space by a function $\lambda(t)$.  We can rewrite the equations of motion in terms of the local functions $s,u$
\begin{align}\label{formal}
\frac{d\lambda}{dt}\partial_{\lambda}v+v\cdot s=-\frac{dP}{d\rho}u+\eta r
\end{align}
In general, we would expect closure relations to be necessary for the evolution of $s$ and $u$ but, for such a simple flow, these are constrained uniquely by the values of $v$ and $\rho$ at each point.  This then becomes an implicit equation for $\dot{\lambda}$.  

In terms of our statistical variables, we see that for each point $x$, the values of $\partial v$, $\partial\partial v$,  $\rho$ and $\partial\rho$ have only one solution for each choice of $v$ obtained at that point.  This means that the probability function $f$ condenses to the form
\begin{align}\label{dist}
f=h(v,p)\delta(s(v,p)-\partial v)\delta(r(v,p)-\rho)\delta(u(v,p)-\partial\rho)\ldots
\end{align}
where the functions $s,r,u$ are assumed known.  The one parameter set of solutions at each point will impose self consistency conditions on such functions since they are not independent of the functions $v(p,t)$ and $\rho(p,t)$.  

We have chosen not to utilize the pressure function as a fundamental variable and instead to keep track of the density $\rho$ with an implicit equation of state $P(\rho)$.  In the case of a constant density fluid we can derive pressure from an elliptic constraint equation of the flow but, for our new statistical variables, some correlation data will almost certainly enter and this becomes intractable for our distribution function in the form $f(p,v,\partial v,\ldots P,\partial P,\ldots)$.  

As the flow gets more complicated we cannot assume that the single valued nature of the variables is preserved so that knowing $h(v,p)$ is sufficient probability to determine the entire solution.  In fact, we know from the case of the Lorenz attractor that the distribution function may not even be quasicontinuous as this grows in complexity.  However, we can be sure that, if we drive the flow uniformly with stationary boundary conditions that the flow will converge to a long time average distribution on some time scale even if it is turbulent.  

Our ``solution'' in eqn.\ \ref{formal} was rather formal and seems to buy us no new information.  Let us now use it as a template to describe more complicated flows.  As in the case of nonlinear differential equations, it seems likely that the local flow data at each point undergo a set of bifurcations as the flow becomes more extreme.  Specifically, the laminar flow leads to oscillations so $v$ and $\rho$ at each point move in a bounded cycle so that the probability $h(v,\rho,p)$ at each point gives an orbit in 4-D space.  This orbit can become very convoluted and never self intersect.  As long as the distribution $h(x,v)$ is such that only one choice of values of $\rho$, $\partial v$ and $\partial\partial v$ can correspond to each $v$, the distribution function in eqn.\ \ref{dist} decomposes and the evolution of the stable distribution(s) is simple.  

When this simple periodicity no longer holds, one can look for higher dimensional correlation functions like $h(v,\rho,\partial v, \partial \rho ,p)$ so that a particular choice of $v,\rho$ at any given point are not sufficient to determine all the required derivatives in the N-S equations uniquely.  Thus, instead of a closure scheme, we are confronted with a pde with boundary conditions where we seek a static solution to a distribution where the dimensionality of the solution may become dependent on the complexity of the flow.  

%The function $h(x,v)$ is a probability distribution at each $x$.  
%
%Our phase space becomes a degenerate 6-D surface at each point.  Furthermore, these functions allow us to apply the N-S equations to find our functions $h,s,v$.  If the value $v$ at $x$ is obtained for only one global profile then we can determine a class of global solutions based on the value of $v$ at one point.  

Observations of turbulent flow give some qualitative observations on the size and strength of vortices so that knowing one component of the velocity at a point puts strong limits on the other directions of flow and its gradients.  Turbulence gives a range of vortex sizes and energies but the observation of scaling laws of Kolmogorov and Richardson \cite{Frisch} suggests some ordering in the energy of vortices based on size.   Because of this and the above bifurcation arguments, we begin with the assumption that the time averaged filling of the local jet spaces for unsteady flows, even turbulent ones, is rather sparse.  Specifically, given a probability function $h(v,\rho,p)$, once $v$ and $\rho$ are specified there is only one local flow profile corresponding to it.  This seems like a large assumption but in the case of the vortex street, it is clearly so.  

Consider such a density function found by a long time observation.  Not just any such function is meaningful.  These are found by averaging over many global flows.  As such, we expect some consistency constraint on such functions.  When we start with $h(v,\rho,p_{o})$ at a particular point $p_{o}$ this should correspond to one global flow for a particular $v,\rho$ there.  This means we should be able to extend our solution outwards from this point by looking for similar $h$ weights in the neighboring points.  Of course, if $h$ is constant in space this is not sufficiently restrictive.  We know that linear shear flow between two parallel plates under high $Re$ numbers is a solution of N-S but not a stable one.  It is reasonable to expect physical solutions will involve $h$ functions of variability in all directions in the space $(v,\rho,p)$.  An immediate problem we face is that restricting $h(v,\rho,p)$ to a constant value in a neighborhood of $p$ does not uniquely fix $v$ and $\rho$ in it.  Therefore, we will need a collection of such functions.  A set that meets this need is the topic of the next section.  

Our goal is to find one (or at least a sparse set of) solutions corresponding to the stationary driving boundary conditions for a flow.  To this end we introduce a time dependence for $h$ as $h(v,\rho,p,t)$ reminiscent of the ensemble approaches in statistical physics.  However, we are not attempting to imply some sort of local, temporal or ensemble averaging of questionable meaning in this case.  We know how N-S evolves the implied solutions and therefore how such a distribution should evolve.  By using some sort of annealing procedure we hope to find stationary physical $h$ functions that give the distribution of the flows over long times.  These distributions will degenerate to particular values at the boundaries or other locations where driving conditions are present.

\section{The Differential Distribution Equations}

We have seen that there is reason to suspect that the statistical distributions from long time averages of a flow with stationary driving conditions will tend to be fairly sparse.  However there is some problem with getting the relations between the spaces at neighboring points to relate to the N-S equations from a single correlated probability function $h(v,\rho,p)$.  Let us begin with the notion that the probability of a given $v_{x}$ at point $x_{o}$ tends to a distribution that we can use to bootstrap our way to correlated probabilities of the $v_{y}$, $v_{z}$ and $\rho$ components.  The field gradients are to be determined by the self consistency condition that the probability functions evaluated at constant value on a neighborhood induce a region of a true solution in time evolution of the system regardless of the initial data.  

To this end introduce the following probability functions at each point: 
\begin{align}
&h_{1}(v_{x},p)\\
&h_{2}(v_{x},v_{y},p)\\
&h_{3}(v_{x},v_{y},v_{z},p)\\
&h_{4}(v_{x},v_{y},v_{z},\rho,p)
\end{align}
To extract the local information about the solution with value $v_{x}^{o}$ at point with coordinates $p_{o}$ we evaluate the value of $h_{1}$ at $(v_{x}^{o}, p_{o})$ and find the values of $v_{x}$ that force $h_{1}$ to be constant in the neighborhood of $p_{o}$.  The dimensionality of the argument implies that there is, generally, one such solution in this neighborhood so that we can expand $v_{x}=v_{x}^{o}+u_{p}\cdot\nabla_{p}v_{x}+\frac{1}{2}u_{p}u_{p}:\nabla_{p}\nabla_{p}v_{x}\ldots=v_{x}^{o}+u_{1}$ where $u_{1}$ is a function of a small $u_{p}$ direction vector defined by $p=p_{o}+u_{p}$.  To linear approximation, the constant value $h_{1}$-sheet is locally defined by 
\begin{align}
\tilde{u}\cdot\nabla_{(v_{x},p)}h_{1}&=(u_{1},u_{p})\cdot\nabla_{(v_{x},p)}h_{1}\\
&=u_{p}\cdot\nabla_{p}h_{1}+u_{1}\partial_{v_{x}}h_{1}=0
\end{align}
so that 
\begin{align}
u_{1}\approx-\frac{u_{p}\cdot\nabla_{p}h_{1}}{\partial_{v_{x}}h_{1}}
\end{align}
and
\begin{align}
v(p+u_{p})\approx v_{o}-\frac{u_{p}\cdot\nabla_{p}h_{1}}{\partial_{v_{x}}h_{1}}
\end{align}

%The second order condition is
%\begin{align}
%\tilde{u}'\tilde{u}':\nabla_{(v_{x},p)}\nabla_{(v_{x},p)}h_{1}&=(u_{2},u_{p})(u_{2},u_{p}):\nabla_{(v_{x},p)}\nabla_{(v_{x},p)}h_{1}\\
%&=u_{p}u_{p}:\nabla_{p}\nabla_{p}h_{1}+u_{2}^{2}\partial_{v_{x}}^{2}h_{1}=0
%\end{align}
%so
%\begin{align}
%u_{2}^{2}=-\frac{u_{p}u_{p}:\nabla_{p}\nabla_{p}h_{1}}{\partial_{v_{x}}^{2}h_{1}}
%\end{align}
However, we need second order information to utilize the viscosity terms in the N-S equations.  The conditions to second order can be found by realizing that $h_{1}(v_{x},p)$ can be expanded as $h_{1}=v_{o}+\tilde{u}\cdot\nabla_{(v_{x},p)}h_{1}+\frac{1}{2}\tilde{u}\tilde{u}:\nabla_{(v_{x},p)}\nabla_{(v_{x},p)}h_{1}\ldots$  Requiring the probability function to be constant locally gives
\begin{align}
u_{p}\cdot\nabla_{p}h_{1}+u_{1}\partial_{v_{x}}h_{1}
+\frac{1}{2}\left( u_{p}u_{p}:\nabla_{p}\nabla_{p}h_{1}+2 u_{1}u_{p}\cdot\nabla_{p}\partial_{v_{x}}h_{1}+u_{1}^{2}\partial_{v_{x}}^{2}h_{1}   \right)=0
\end{align}
or
\begin{align}
\left(\frac{1}{2}\partial_{v_{x}}^{2}h_{1}\right)u_{1}^{2} +\left(\partial_{v_{x}}h_{1}+u_{p}\cdot\nabla_{p}\partial_{v_{x}}h_{1}\right)u_{1}+
\left(u_{p}\cdot\nabla_{p}h_{1}
+\frac{1}{2} u_{p}u_{p}:\nabla_{p}\nabla_{p}h_{1}\right)=0
\end{align} 
Choosing the branch of solutions connected to the above linear one we have
\begin{align}
u_{1}=-u_{i}\frac{\nabla_{i}h_{1}}{\partial_{v_{x}}h_{1}}
+u_{i}u_{j}\frac{1}{2}\left(-\frac{\nabla_{i}\nabla_{j}h_{1}}{\partial_{v_{x}}h_{1}}
+\frac{\nabla_{i}(\partial_{v_{x}}h_{1})\nabla_{j}h_{1}}{(\partial_{v_{x}}h_{1})^{2}}
+\frac{\nabla_{j}(\partial_{v_{x}}h_{1})\nabla_{i}h_{1}}{(\partial_{v_{x}}h_{1})^{2}}
-\frac{\partial_{v_{x}}h_{1}\nabla_{i}h\nabla_{j}h_{1}}{(\partial_{v_{x}}h_{1})^{3}}     \right)
\end{align}
from which we determine that
\begin{align}
\nabla_{i}v_{x}=-\frac{\nabla_{i}h_{1}}{\partial_{v_{x}}h_{1}}
\end{align}
and
\begin{align}
\nabla_{i}\nabla_{j}v_{x}=\left(-\frac{\nabla_{i}\nabla_{j}h_{1}}{\partial_{v_{x}}h_{1}}
+\frac{\nabla_{i}(\partial_{v_{x}}h_{1})\nabla_{j}h_{1}}{(\partial_{v_{x}}h_{1})^{2}}
+\frac{\nabla_{j}(\partial_{v_{x}}h_{1})\nabla_{i}h_{1}}{(\partial_{v_{x}}h_{1})^{2}}
-\frac{\partial_{v_{x}}h_{1}\nabla_{i}h\nabla_{j}h_{1}}{(\partial_{v_{x}}h_{1})^{3}}     \right)
\end{align}

%\label{a}
Substitution in the N-S equations gives us equations for $a_{x}=\partial_{t}v_{x}$ in terms of $v$, $\nabla_{i}v_{x}$, $\nabla_{i}P(\rho)$, and $\nabla^{2}v_{x}$.  This still leaves the missing parts $v_{y}$, $v_{z}$, $\rho$ and $\nabla_{i}\rho$.  Interestingly, the N-S equations are linear in the velocities in the viscous term but not in the advective term $v\cdot\nabla v$.  We now note that we have a local second order representation for $v_{x}$ we can use $h_{2}(v_{x},v_{y},p)$ and substitute this result to give $\tilde{h}_{2}(v_{y},p)$.  By iterating the above process we obtain equations for $v_{y}$.  Similarly, $h_{3}$ and $h_{4}$ give the missing $v_{z}$ and $\rho$ information.  The equations for $a_{y}$ and $a_{z}$ arise in the same fashion.  The conservation law specifies $\partial_{t}\rho=-\nabla_{p}\cdot(\rho v)$.  These are all in terms of our probability functions, $h_{k}$, and gradients of them.  

%where we define $u(\hat{w},p)$ by $(\hat{w},u)\cdot\nabla_{(p,v)}h_{1}=0$ and $\hat{w}$ is the unit vector in the $p-p_{o}$ direction.  By evaluating $v_{x}$ to second order in a neighborhood of $x_{o}$ we can find the derivatives of it necessary for the N-S equations.  

%The evolution of $\rho$ is given by $\partial_{t}\rho=-\nabla_{p}\cdot(\rho v)$.  Similarly, we then find the local derivatives of $\rho$ about $p_{o}$ from $h_{4}$.  Notice that we needed to have the probability functions with ever increasing numbers of velocity variables starting with only $v_{x}$ to be able to get the local patch solution from them.  

The above repeated iteration allows us to bootstrap our way to the second order derivatives of all the velocity functions $(v_{x},v_{y},v_{z})$ and $\rho$ while allowing correlations among them.  Our goal is not now utilize these the accelerations $a_{j}$ from the N-S equations to specify the time derivatives of the probability functions as a closed system.  To accomplish this for $h_{1}$ we need to transfer the information from each solution patch and its evolution to obtain the time dependence of the probability function $h_{1}(p,v)$.  The N-S solution tells us how $v$ is locally changing through $a=\partial_{t}v$.  To map this to an action on our probability function we view this as a push-forwards of the distribution function $\tilde{h}_{1}(v_{x})=h_{1}|_{p}$ at each point.  Since $\tilde{h}_{1}$ is a density the transformation $v\rightarrow v+a\delta t$ is induced by the shift in the coordinates and a factor of the measure of the transformation.  The Jacobian of the transformation is $J=\det(\partial_{v_{x}}(v_{x}+a_{x}\delta t))=1+\partial_{v_{x}}(a_{x}\delta t)$ where the location $p$ is fixed.  The function $a(v,p)$ is a function of second spatial derivatives of $h_{1}$ but now we interpret it as vector field in the local space labeled by $v_{x}$ so the derivatives wrt $v_{x}$ require no higher order derivatives of $h_{1}$.  This measure factor ensures that our normalization condition is preserved and $\int dv \dot{h}_{1}=0$.  The evolution of $h_{1}$ is then given by the  infinitesimal evolution of the distribution $h_{1}(v_{x}(t),p)=h_{1}(v_{x}+a_{x}\delta t,p)\cdot J(v_{x}+a_{x}\delta t,v_{x},p)$ 
\begin{align}
\dot{h}_{1}(v_{x},p,t)&=(\partial_{v_{x}}h_{1})(\partial_{t}v_{x})+h_{1}(v_{x},p)\cdot \frac{J-1}{\delta t}\\
&= (\partial_{v_{x}}h_{1})\cdot  a_{x}+h_{1} \cdot\partial_{v_{x}} a_{x}\\
&= \partial_{v_{x}}(a_{x} h_{1})
\end{align}
where $a$ is given indirectly by the N-S equations through $\partial_{t}v$.  We have taken some liberty with the single derivative ``dot'' symbol here not to imply $h_{1}$ is a function of one variable but to indicate that the function $h_{1}$ lives on the fixed coordinate space $(v_{x},p,t)$ so none of the arguments evolve hence no chain rule factors arise.  

\begin{figure}[!ht]
   \centering
   \includegraphics[width=4in,trim=0mm 0mm 0mm 0mm,clip]{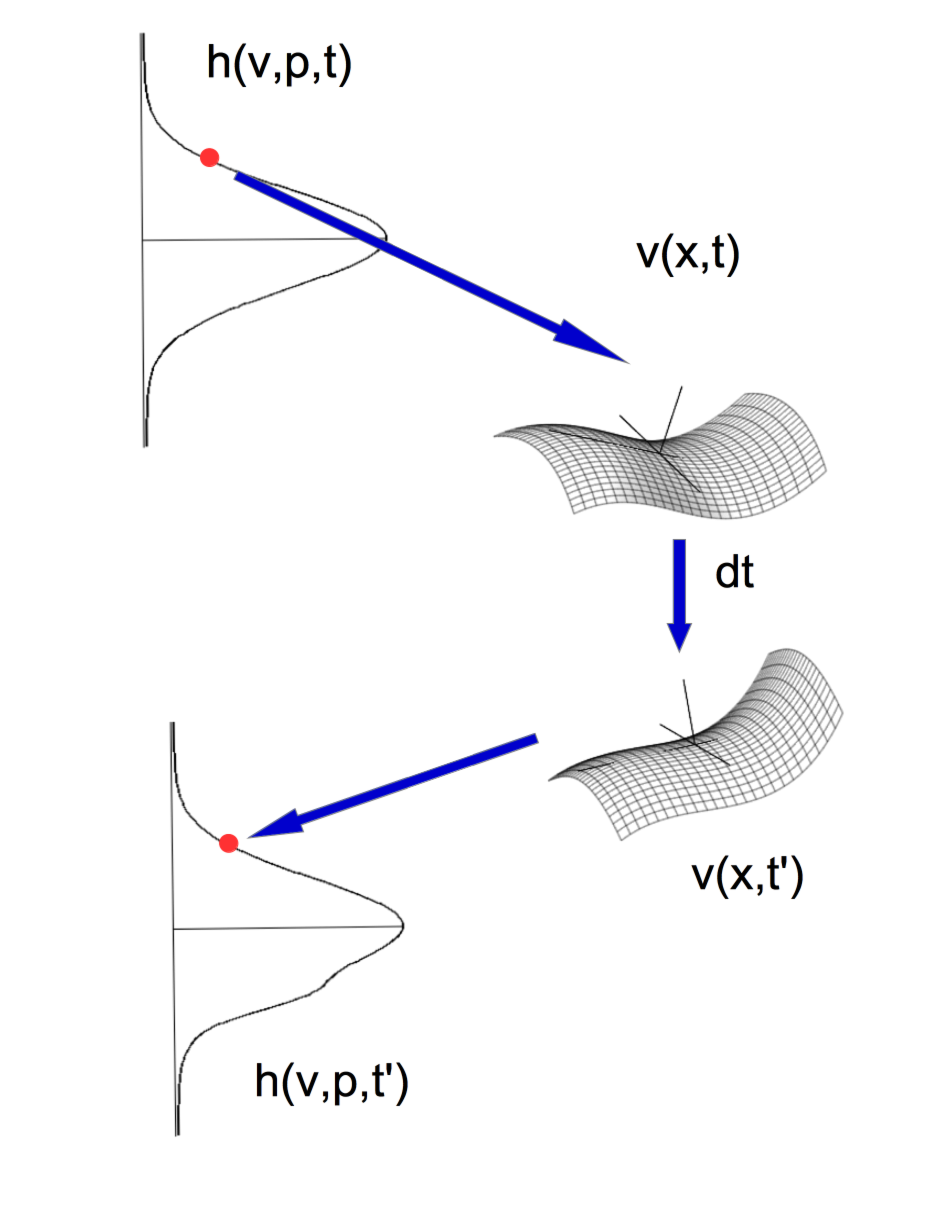} 
   \caption{Hydrodynamic probability data at $p$ yields a unique corresponding local velocity field about $p$ that is evolved by N-S then mapped to back to a flow on the local probability space.}
   \label{flow}
\end{figure}

We can iteratively apply this argument to get equations for the time derivatives of each of the distribution functions to give a set of four coupled partial differential equations.  For the case of $h_{2}$ we have 
\begin{align}
\dot{h}_{2}(v_{x},v_{y},p,t)&=(\partial_{v_{x}}h_{2})\cdot a_{x}+(\partial_{v_{y}}h_{2}) \cdot a_{y}+h_{2}(v_{x},v_{y},p)\cdot \frac{J-1}{\delta t}\\
\end{align}
where the Jacobian of the transformation is $J=\det(\partial_{v_{i}}(v_{j}+a_{j}\delta t))$ is the determinant of a 2x2 matrix in $x$ and $y$ components.  The results for $\dot{h}_{3}$ and $\dot{h}_{4}$ are similar, though of increasing complexity due to the iterations.  The calculational flow for these repeated steps can be summarized in fig.\ \ref{flow}

The physical solution is the case where the time derivatives vanish as we have argued that this is the limiting time average of the stationarily driven flow.  Certainly such a limit exists.  It is not clear that it is unique or that our solution could not have combined multiple distinct results.  In nonlinear dynamics of ODEs one can have multiple separate attractors.  Certainly we can have distinct unstable flows as in the laminar case.  In the case of turbulence or randomly perturbed situations, it is unknown if such a situation can exist.  For now we assume that our flows have no such complications.  

A natural approach to solving such a system is to begin with some sufficiently nontrivial initial data so that unique local solution patches can be inferred from our constant probability conditions then to ``anneal'' the distributions by removing probability from the rapidly changing parts of each probability function at each $p$ and renormalizing so that the norm of each is preserved.  

\section{Other Systems}
The motivations for this approach to hydrodynamic flow came from investigations into thermal systems, specifically, the thermal quantum state.  Although physicists have grown more comfortable with a deterministic many body wavefunction as the universal descriptor for all time in nature, the extension of this to thermodynamics has been especially troublesome.  The main problems with quantum mechanics as a deterministic system stem from discrete measurement actions.  This is presumed to be solved by decoherence by many, often by updating the very definition of decoherence to meet new challenges and problems.  In the case of quantum statistical mechanics, ensembles are used rather formally\cite{Tolman}.\footnote{The could sometimes be construed as recklessly since the main justification often for their use is the selective collection of a posteriori successes and historical inertia.}  The Boltzmann counting is used among eigenstates near a given energy to give density of states function from which all thermodynamic properties are derived \cite{Chafin2}.  This is problematic in the case of discrete states where such a quasicontinuous averaging is ambiguous but more so in the nonequilibrium case.  One can make even more severe objections to the fact that one should be able to have constant energy states that are made from states nowhere near the same eigenstate energies.  In the case of classical kinetics, the constant energy surface constrains the gas in phase space but such a space in the quantum Hilbert space is far larger than that limited by the quantum ensemble.  

To overcome this problem and find a description that is more compatible with the actual many body wavefunction of a particular system that is overwhelmingly likely to give statistical behavior, in the spirit of classical kinetics, we can attempt a time averaged state in many body space $\tilde{X}=\{x_{1},x_{2}\ldots x_{N}\}$ we can give two functions $h_{1}(\Re[\Psi(\tilde{X})])$ and $h_{2}(\Re[\Psi(\tilde{X})],\Im[\Psi(\tilde{X})])$. Equivalently, one could use the many body density $\rho(\tilde{X})$ and current $J(\tilde{X})$ with the proviso that the current be finite and irrotational except at $3N-2$ dimensional subspaces where the density vanishes.  The problem of deriving quantum kinetics from first principles persists despite the frequent usefulness of the Kubo formula.  It is clear that one must have a spread in eigenstate energies to get any fluctuations.  On the other hand, if one needs a particular distribution of energies then the property of self equilibration of a system, as we generally see in nature, would not be present.  A local statistical distribution made from the long time averages might provide some universality for a broad class of physical initial data.  

In a more classical direction, hydrodynamics of classical gases can be derived to lowest order from the Maxwell-Boltzmann equations themselves \cite{Jeans}.  However, attempts at higher order corrections typically lead to divergences.  The problem seems likely to be in the local averaging we use to define our thermodynamic variables.  The density and velocity fields require some space and/or time averaging to get smoothed results.  When mean-free-path scale effects become important such averaging is no longer valid.  Fluctuations on the small scale are suspected to be important \cite{Romatschke} and this is when it is said that hydro ``breaks.''  

A possible approach that does not rely on ensembles or such regional averaging is to give nonlocal variables that measure, at each instance, the weighted distribution of particle neighbor location about every point (independent of wether there is a particle at the point or not).  By extending to many particle separations, one can get a rather smooth distribution of quasilocal information that is associated with that point.  By finding a self consistency condition for such measures one might evolve statistical measure of such distributions using annealing methods to reach a time independent result corresponding to the long time average of the classical kinetic system.  

\section{Conclusions}
The use of ensembles in physics have many conceptual difficulties and are introduced ``formally'' so often that these problems are now rarely known well even to practitioners.  This collection of probability functions is introduced in such a way as to avoid these.  The only physical meaningful solutions are the final time independent solutions with the time dependent ones only transitional tools to arrive at them.  The boundary conditions are specified on the distribution function by no-slip conditions so that the distribution degenerates to a single value of $v$ there.  The pressure is not necessarily constant but the perpendicular acceleration of $v$ should remain zero which induces an indirect constraint on the density there.  
Even if this method should solve many problems of turbulent flow they still have obvious limitations.  We often care about transitions to turbulence and cases where the external forces and boundaries are not constant.  One might be tempted to utilize these probability functions in such cases but care here is essential.  The meaning of such a distribution is not clear even if it should produce a set of desired boundary stresses and internal dynamics and energy distribution though some relaxation.  

The use of statistical methods in physics has been a powerful tool since Maxwell introduced them for the theory of gases.  In hydrodynamics, flows can have energy distributed on so many scales that numerical simulations rapidly run into speed and data storage limits.  An analytic approach to turbulence based on statistical data has evident advantages even if it is only a stepping stone to more condensed numerical methods.  In the case here we have a reasonably compact set of functions with no arbitrary freedom in the evolution equations.  The biggest assumption is in the sparseness of the distribution so we can use local statistical data to match with a local hydrodynamic patch of data to evolve and induce a closure in the evolution of the distribution functions themselves.  It is quite possible that this condition fails at some level of turbulence.  Extensions of this method could involve some additional local differential data in the distribution functions to obtain the desired sparseness or a kind of statistical ``connection'' function that joins such local data and also requires some self consistent evolution.

%To apply the information in the N-S equations we can view the evolution of $v_{1}$ at each point as a flow that preserved the local probability density $\int dv h_{1}=1$.  The N-S equations tell us how $v$ evolves.  The since the probability is a measure we need a Jacobian weight  due to convergence and divergence of these flows but since $h_{1}|_{p_{o}}$ is a function of only one variable this is $D=\partial_{v_{x}}h_{1}$.
%The evolution is given by
%\begin{align}
%\partial_{t} h_{1}=\partial_{v_{x}}h_{1}\partial_{t}v_{x}
%\end{align}
%where the time derivatives of $v_{x}$ are to be given by N-S.  Of course, N-S has derivatives with respect to the other variables as well so this is not a closed solution yet.  
%
%%vdot induces a flow.  use convective derivative.  
%$\dot{v}\nabla_{v} h$
%
%$h(v(t),x)$  $\det(\partial_{v}\dot{v})$
%
%$\det(\partial_{v}h)$

%Show the order of v1, v2... in these nested equations does not matter.  The choice of v1 in h1 should give no anisotropy versus v2 in h1.  

\end{document}